\documentstyle[12pt] {article}
\newcommand{\beq}{\begin{equation}}
\newcommand{\eeq}{\end{equation}}
\newcommand{\beqa}{\begin{eqnarray}}
\newcommand{\eeqa}{\end{eqnarray}}
\newcommand{\ba}{\begin{array}}
\newcommand{\ea}{\end{array}}

\begin{document}

\begin{center}
{\large \bf Quantum Chaos in a Yang--Mills--Higgs System}
\end{center}

\vskip 1. truecm

\begin{center}
{\bf Luca Salasnich}
\footnote{E--Mail: salasnich@padova.infn.it}
\vskip 0.5 truecm
Dipartimento di Matematica Pura ed Applicata,\\
Universit\`a di Padova, Via Belzoni 7, I 35131 Padova, Italy\\
and\\
Istituto Nazionale di Fisica Nucleare, Sezione di Padova, \\
Via Marzolo 8, I 35131 Padova, Italy \\
\end{center}

\vskip 1. truecm

\begin{center}
{\bf Abstract}
\end{center}

\vskip 0.5 truecm
\par
We study the energy fluctuations 
of a spatially homogeneous SU(2) Yang--Mills--Higgs system. 
In particular, we analyze the nearest--neighbour spacing 
distribution which shows 
a Wigner--Poisson transition by increasing the value of the Higgs 
field in the vacuum. This transition is a clear quantum signature 
of the classical chaos--order transition of the system.  

\vskip 0.5 truecm
\begin{center}
To be published in Modern Physics Letters A
\end{center}
 
\newpage

\section{Introduction}
\par
In the last years there has been much interest in classical chaos 
in field theories. It is now well known that the spatially uniform limits 
of scalar electrodynamics and Yang--Mills theory exhibit classical 
chaotic motion$^{1)-8)}$. On the other hand, in field theories, 
less attention has been paid to {\it quantum chaos}, 
i.e. the study of properties of quantum systems 
which are classically chaotic$^{9)}$. 
\par
The energy fluctuation properties of systems with underlying 
classical chaotic behaviour and time--reversal symmetry agree with 
the predictions of the Gaussian Orthogonal Ensemble (GOE) of 
random matrix theory, whereas quantum analogs of classically 
integrable systems display the characteristics 
of the Poisson statistics$^{9)-12)}$. 
Some results in this direction for field theories have been obtained by 
Halasz and Verbaarschot: they studied the QCD lattice spectra 
for staggered fermions and its connection to random matrix theory$^{13)}$. 
\par
In this paper we study quantum chaos in a field--theory schematic model. 
We analyze the energy fluctuation properties 
of the spatially homogeneous SU(2) Yang--Mills--Higgs (YMH) system 
(see Ref. 1--4). We show that these fluctuations give a clear quantum signature 
of the classical chaos--order transition of the system. 
\par
The Lagrangian density of the SU(2) YMH system$^{14)}$ is given by 
\beq
L={1\over 2}(D_{\mu}\phi )^+(D^{\mu}\phi ) -V(\phi ) 
-{1\over 4}F_{\mu \nu}^{a}F^{\mu \nu a} \; ,
\eeq
where
\beq
(D_{\mu}\phi )=\partial_{\mu}\phi - i g A_{\mu}^b T^b\phi 
\; ,
\eeq
\beq
F_{\mu \nu}^{a}=\partial_{\mu}A_{\nu}^{a}-\partial_{\nu}A_{\mu}^{a}+
g\epsilon^{abc}A_{\mu}^{b}A_{\nu}^{c} \; ,
\eeq
with $T^b=\sigma^b/2$, $b=1,2,3$, generators of the SU(2) algebra, 
and where the potential of the scalar field (the Higgs field) is
\beq
V(\phi )=\mu^2 |\phi|^2 + \lambda |\phi|^4 \; .
\eeq
We work in the (2+1)--dimensional Minkowski space ($\mu =0,1,2$) and 
choose spatially homogeneous Yang--Mills and the Higgs fields
\beq
\partial_i A^a_{\mu} = \partial_i \phi = 0 \; , \;\;\;\; i=1,2
\eeq
i.e. we consider the system in the region in which space fluctuations of 
fields are negligible compared to their time fluctuations. 
\par
In the gauge $A^a_0=0$ and using the real triplet representation for the 
Higgs field we obtain
$$
L={\dot{\vec \phi}}^2 +
{1\over 2}({\dot {\vec A}}_1^2+{\dot {\vec A}}_2^2) 
-g^2 [{1\over 2}{\vec A}_1^2 {\vec A}_2^2 
-{1\over 2} ({\vec A}_1 \cdot {\vec A}_2)^2+
$$
\beq
+({\vec A}_1^2+{\vec A}_2^2){\vec \phi}^2 
-({\vec A}_1\cdot {\vec \phi})^2 -({\vec A}_2 \cdot {\vec \phi})^2 ]
-V( {\vec \phi} ) \; ,
\eeq
where ${\vec \phi}=(\phi^1,\phi^2,\phi^3)$, 
${\vec A}_1=(A_1^1,A_1^2,A_1^3)$ and ${\vec A}_2=(A_2^1,A_2^2,A_2^3)$. 
\par
When $\mu^2 >0$ the potential $V$ has a minimum at $|{\vec \phi}|=0$, 
but for $\mu^2 <0$ the minimum is at 
$$
|{\vec \phi}_0|=\sqrt{-\mu^2\over 4\lambda }=v \; ,
$$
which is the non zero Higgs vacuum. This vacuum is degenerate 
and after spontaneous symmetry breaking the physical vacuum can be 
chosen ${\vec \phi}_0 =(0,0,v)$. If $A_1^1=q_1$, $A_2^2=q_2$ 
and the other components of the Yang--Mills fields are zero, 
in the Higgs vacuum the Hamiltonian of the system reads 
\beq
H={1\over 2}(p_1^2+p_2^2)
+g^2v^2(q_1^2+q_2^2)+{1\over 2}g^2 q_1^2 q_2^2 \; ,
\eeq
where $p_1={\dot q_1}$ and $p_2={\dot q_2}$. Here $w^2=2 g^2v^2$ is the 
mass term of the Yang--Mills fields. This YMH Hamiltonian is 
a toy model for classical non--linear dynamics, with the attractive feature 
that the model emerges from particle physics. In the next sections we 
analyze first the classical chaos--order transition of the YMH system 
and then its connection to the quantal fluctuations of the energy levels. 

\section{Classical chaos--order transition}

\par
A classical chaos--order transition for the YMH system 
has been observed previously by different authors: 
Savvidy used the Chirikov'criterion$^{1)}$, Kawabe and Ohta studied 
the Lyapunov exponents$^{3)}$ and Salasnich analyzed the 
quantal overlapping resonances$^{4)}$. 
In this paper we study the chaotic behaviour of this YMH system by using 
the Gaussian curvature criterion of the 
potential energy$^{16)}$ and the Poincar\`e Sections$^{17)}$. 
\par
At low energy the motion near 
the minimum of the potential 
\beq 
V(q_1,q_2)=g^2 v^2 (q_1^2+q_2^2)+{1\over 2} g^2 q_1^2 q_2^2 \; ,
\eeq 
where the Gaussian curvature is positive, is periodic or quasiperiodic and is 
separated from the instability region by a line of zero curvature; 
if the energy is increased, the system will be for some initial conditions 
in a region of negative curvature, where the motion is chaotic. 
According to this scenario, the energy $E_c$ of chaos--order transition 
is equal to the minimum value of the line of zero gaussian 
curvature $K(q_1 ,q_2 )$ on the potential--energy surface. 
For our potential the gaussian curvature vanishes at the points 
that satisfy the equation
\beq
{\partial^2 V \over \partial q_1^2}
{\partial^2 V \over \partial q_2^2}-
({\partial^2 V \over \partial q_1 \partial q_2})^2=
(2g^2v^2 +g^2 q_2^2)(2g^2v^2+g^2q_1^2)-4g^4 q_1^2q_2^2=0 \; .
\eeq
It is easy to show that the minimal energy on the 
zero--curvature line is given by:
\beq
E_c=V_{min}(K=0,\bar{q_1})=6 g^2 v^4 \; , 
\eeq
and by inverting this equation 
we obtain $v_c=(E /6g^2)^{1/4}$. We conclude that 
there is a order--chaos transition by increasing the energy $E$ 
of the system and a chaos--order transition by increasing 
the value $v$ of the Higgs field in the vacuum (see also Ref. 2). Thus, 
there is only one transition regulated by the unique parameter $E/(g^2v^4)$. 
\par
It is important to point out that 
{\it in general} the curvature 
criterion guarantees only a {\it local instability}$^{16)}$ 
and should therefore 
be combined with the Poincar\`e sections$^{17)}$ (see Ref. 18). 
The classical equations of motion of the YMH system are
\beq
{\dot q_1}=p_1 \; , \;\;\;\; 
{\dot q_2}=p_2 \; , \;\;\;\; 
{\dot p_1}=-2g^2v^2 q_1 - g^2 q_1 q_2^2 \; , \;\;\;\; 
{\dot p_2}=-2g^2v^2 q_2 - g^2 q_1^2 q_2 \; . 
\eeq
We use a fourth--order Runge--Kutta 
method$^{19)}$ to compute the classical trajectories. 
The conservation of energy restricts any trajectory of the four--dimensional 
phase space to a three--dimensional energy shell. At a particular energy 
the restriction $q_1=0$ defines a two--dimensional surface in 
the phase space, which is called Poincar\`e section. 
Each time a particular trajectory passes through the surface 
a point is plotted at the position of intersection $(q_2,p_2)$. 
We employ a first--order interpolation process to reduce the inaccuracies 
due to the use of a finite step length$^{17)}$. 
\par
In Figure 1 we plot the Poincar\`e sections for different values of 
the Higgs vacuum $v$ but with the same energy $E$ and 
interaction $g$. Chaotic regions on the surface of section 
are characterized by a set of randomly distributed points 
and regular regions by dotted or solid curves. 
The pictures show that the parameter $v$ plays 
an important role: for large values it makes the system regular. 
In fact, if we increase the harmonic part of the YMH potential 
the effect of the nonlinear term becomes less important. 
These numerical calculations confirm the analytical 
predictions of the curvature criterion: with $E =10$ and 
$g=1$ we get the critical value of the onset of chaos 
$v_c=(E /6g^2)^{1/4}\simeq 1.14$, 
in very good agreement with the Poincar\`e sections. 

\section{Quantum signature of the chaos--order transition}
\par
In quantum mechanics the generalized coordinates of the YMH system 
satisfy the usual commutation rules $[{\hat q}_k,{\hat p}_l]=i\delta_{kl}$, 
with $k,l=1,2$. Introducing the creation and destruction operators
\beq
{\hat a}_k=\sqrt{\omega \over 2}{\hat q}_k + 
i \sqrt{1\over 2\omega}{\hat p}_k \; ,
\;\;\;\;
{\hat a}_k^+ = \sqrt{\omega \over 2}{\hat q}_k - 
i \sqrt{1\over 2\omega}{\hat p}_k \; ,
\eeq
the quantum YMH Hamiltonian can be written$^{15)}$ 
\beq
{\hat H}={\hat H}_0 + {1\over 2} g^2 {\hat V} \; ,
\eeq
where
\beq
{\hat H}_0= \omega ({\hat a}_1^+ {\hat a}_1 + {\hat a}_2^+ {\hat a}_2 + 1) \; ,
\eeq
\beq
{\hat V}= {1 \over 4 \omega^2} ({\hat a}_1 +{\hat a}_1^+)^2 
({\hat a}_2 +{\hat a}_2^+)^2 \; ,
\eeq
with $\omega^2 = 2 g^2 v^2$ and $[{\hat a}_k,{\hat a}_l^+] = \delta_{kl}$, 
$k,l=1,2$. 
\par
The most used quantity to study the local fluctuations of the energy levels 
is the spectral statistics $P(s)$. $P(s)$ is 
the distribution of nearest--neighbour spacings 
$s_i=({\tilde E}_{i+1}-{\tilde E}_i)$ 
of the unfolded levels ${\tilde E}_i$. 
It is obtained by accumulating the number of spacings that lie within 
the bin $(s,s+\Delta s)$ and then normalizing $P(s)$ to unity$^{9)-12)}$. 
\par
For quantum systems whose classical analogs are integrable, 
$P(s)$ is expected to follow the Poisson limit, i.e. 
$P(s)=\exp{(-s)}$. On the other hand, 
quantal analogs of chaotic systems exhibit the spectral properties of 
GOE with $P(s)= (\pi / 2) s \exp{(-{\pi \over 4}s^2)}$, which is the 
so--called Wigner distribution$^{9)-12)}$. 
The distribution $P(s)$ is the best spectral statistics to analyze 
shorter series of energy levels and 
the intermediate regions between order and chaos. 
\par
Seligman, Verbaarschot and Zirnbauer$^{20)}$ analyzed 
a class of two--dimensional anharmonic oscillators with 
polynomial perturbation by using the Brody distribution$^{21)}$ 
\beq
P(s,\omega)=\alpha (\omega +1) s^{\omega} \exp{(-\alpha s^{\omega+1})} \; ,
\eeq
with 
\beq
\alpha = \big( \Gamma [{\omega +2\over \omega+1}] \big)^{\omega +1} \; .
\eeq
This distribution interpolates between the Poisson distribution ($\omega =0$) 
of integrable systems and the Wigner distribution ($\omega =1$) of 
chaotic ones, and thus the parameter $\omega$ can be used as a simple 
quantitative measure of the degree of chaoticity. 
\par
We compute the energy levels $\{ E_i \}$ with 
a numerical diagonalization of the truncated matrix of the quantum 
YMH Hamiltonian  in the basis of the harmonic oscillators$^{22)}$. 
If $|n_1 n_2>$ is the basis of the occupation numbers of the two 
harmonic oscillators, the matrix elements are
\beq
<n_{1}^{'}n_{2}^{'}|{\hat H}_0|n_{1}n_{2}>= \omega (n_1+n_2+1) 
\delta_{n_{1}^{'}n_{1}} \delta_{n_{2}^{'}n_{2}} \; ,
\eeq
and
$$
<n_{1}^{'}n_{2}^{'}|{\hat V}|n_{1}n_{2}>=
{1 \over 4 \omega^2}
[\sqrt{n_{1}(n_{1}-1)} \delta_{n^{'}_{1}n_{1}-2}
+\sqrt{(n_{1}+1)(n_{1}+2)}\delta_{n^{'}_{1}n_{1}+2}+
(2n_{1}+1)\delta_{n^{'}_{1}n_{1}}]\times 
$$
\beq
\times[\sqrt{n_2 (n_2-1)}\delta_{n^{'}_2 n_2-2}+ \sqrt{(n_2+1)(n_2+2)}
\delta_{n^{'}_2 n_2+2}+ (2n_2+1)\delta_{n^{'}_2 n_2}] \; .
\eeq
The symmetry of the potential enables us to split 
the Hamiltonian matrix into 4 sub--matrices 
reducing the computer storage required. These sub--matrices are related 
to the parity of the two occupation numbers $n_1$ and $n_2$: 
even--even, odd--odd, even--odd, odd--even. 
The numerical energy levels depend on the dimension of the truncated matrix: 
we compute the numerical levels in double precision 
increasing the matrix dimension until the first 100 levels converge 
within $8$ digits (matrix dimension $1156\times 1156$)$^{22),23)}$. 
\par
We use the first $100$ energy levels of the 4 sub--matrices 
to calculate the $P(s)$ distribution. 
In order to remove the secular variation of the level density as a function 
of the energy $E$, for each value of the coupling constant the 
corresponding spectrum is mapped, by a numerical procedure described in 
Ref. 24, into one which has a constant level density: 
$\{ E_i \} \to \{ {\tilde E_i} \}$ (unfolding procedure). 
We use the following standard procedure to avoid mixing 
between states of different symmetry classes: 
1) the diagonalization is performed for each sub-matrix 
(first 100 levels for each sub-matrix); 
2) the unfolding is done for each sub-matrix; 
3) the spacings are calculated for each sub-matrix; 
4) the spacings of the 4 sub-matrices are 
accumulated to plot the P(s) distribution. 
\par
In Figure 2 we plot the $P(s)$ distribution 
for different values of the parameter $v$. 
The figure shows a Wigner--Poisson transition by increasing the value $v$ 
of the Higgs field in the vacuum. 
By using the P(s) distribution and the Brody function 
it is possible to give a quantitative measure 
of the degree of quantal chaoticity of the system. 
Our numerical calculations show clearly the quantum 
chaos--order transition and its connection to the classical one.  

\section{Conclusions}
\par
The chaotic behaviour of an 
homogenous YMH system has been studied both in classical and quantum 
mechanics. The Gaussian curvature criterion and the Poincar\`e 
sections show that the chaotic behaviour 
is regulated by the unique parameter $E/(g^2v^4)$. 
The YMH system has a order--chaos transition by 
increasing the energy $E$ and a chaos--order 
transition by increasing the value $v$ of the Higgs field in the vacuum. 
\par
The nearest--neighbour spacing distribution of the energy 
levels confirms with great accuracy 
the classical chaos--order transition of the YMH system. 
In particular, the Brody function shows a Wigner--Poisson transition 
for the $P(s)$ distribution in correspondence to the classical 
chaos--order transition. 
\par 
We observe that, as stressed previously, our YMH system is a toy model 
but it is very useful because it is possible 
to compare classical to quantum chaos. 
In the future will be important to study classical 
and quantum chaos in more realistic field theories. 

\section*{Acknowledgments}

The author is grateful to G. Benettin, V.R. Manfredi, M. Robnik 
and A. Vicini for stimulating discussions. 

\newpage

\parindent=0.pt
\section*{Figure Captions}
\vspace{0.6 cm}

{\bf Figure 1}: The Poincar\`e sections of the model. From the top: 
$v=1$, $v=1.1$ and $v=1.2$. Energy $E = 10$ and interaction $g=1$. 

{\bf Figure 2}: $P(s)$ distribution. From the top: 
$v=1$ ($\omega=0.92$), $v=1.1$ ($\omega =0.34$) and $v=1.2$ ($\omega =0.01$), 
where $\omega$ is the Brody parameter. First 100 energy levels 
and interaction $g=1$. The dotted, dashed and solid curves stand 
for Wigner, Poisson and Brody distributions, respectively. 

\newpage

\section*{References}

\begin{description}

\item{\ 1.} G.K. Savvidy, Nucl. Phys. {\bf B 246}, 302 (1984).

\item{\ 2.} A. Gorski, Acta Phys. Pol. {\bf B 15}, 465 (1984).

\item{\ 3.} T. Kawabe and S. Ohta, Phys. Rev. {\bf D 44}, 1274 (1991).

\item{\ 4.} L. Salasnich, Phys. Rev. {\bf D 52}, 6189 (1995). 

\item{\ 5.} T. Kawabe, Phys. Lett. {\bf B 343}, 254 (1995).

\item{\ 6.} L. Salasnich, Mod. Phys. Lett. {\bf A 10}, 3119 (1995).

\item{\ 7.} J. Segar and M.S. Sriram, Phys. Rev. {\bf D 53}, 3976 (1996).

\item{\ 8.} S.G. Matinyan and B. Muller, 
Phys. Rev. Lett. {\bf 78}, 2515 (1997). 

\item{\ 9.} M.C. Gutzwiller, {\it Chaos in Classical and Quantum Mechanics} 
(Springer, Berlin, 1990).

\item{\ 10.} A.M. Ozorio de Almeida, {\it Hamiltonian Systems: Chaos and 
Quantization} (Cambridge University Press, Cambridge, 1990).

\item{\ 11.} K. Nakamura, {\it Quantum Chaos} 
(Cambridge Nonlinear Science Series, Cambridge, 1993).

\item{\ 12.} G. Casati and B.V. Chirikov, {\it Quantum Chaos} 
(Cambridge University Press, Cambridge, 1995).

\item{\ 13.} M.A. Halasz and 
J.J.M. Verbaarschot, Phys. Rev. Lett. {\bf 74}, 3920 (1995).

\item{\ 14.} C. Itzykson and J.B. Zuber, 
{\it Quantum Field Theory} (McGraw--Hill, New York, 1985).

\item{\ 15.} G.K. Savvidy, Phys. Lett. {\bf B 159}, 325 (1985).

\item{\ 16.} M. Toda, Phys. Lett. {\bf A 48}, 335 (1974).

\item{\ 17.} M. Henon, Physica {\bf D 5}, 412 (1982).

\item{\ 18.} G. Benettin, R. Brambilla and 
L. Galgani, Physica {\bf A 87}, 381 (1977).

\item{\ 19.} Subroutine D02BAF, The NAG Fortran Library, Mark 14  
(NAG Ltd, Oxford, 1990).

\item{\ 20.} T.H. Seligman, J.J.M. Verbaarschot and M.R. Zirnbauer, 
Phys. Rev. lett. {\bf 53}, 215 (1984).

\item{\ 21.} T.A. Brody, Lett. Nuovo Cimento {\bf 7}, 482 (1973).

\item{\ 22.} S. Graffi, V.R. Manfredi and L. Salasnich, 
Mod. Phys. Lett. {\bf B 9}, 747 (1995).

\item{\ 23.} Subroutine F02AAF, The NAG Fortran Library, Mark 14 
(NAG Ltd, Oxford, 1990).

\item{\ 24.} V.R. Manfredi, Lett. Nuovo Cimento {\bf 40}, 135 (1984). 

\end{description}

\end{document}